\documentclass[11pt,a4paper]{article}
\usepackage{graphicx}

\renewcommand{\dj}{\hbox{d\hskip-1.1ex{\raise0.640ex\hbox{--}}\skip 0.70ex}}
\newcommand{\calQ}{\mathcal{Q}}

\newcommand{\calL}{\mathcal{L}}
\newcommand{\bra}[1]{\langle #1 |}
\newcommand{\ket}[1]{| #1 \rangle}

\begin{document}

\thispagestyle{empty}
\begin{flushright}
\end{flushright}



\begin{center}

  \begin{Large}

  \begin{bf}


Unexpectedly small empirical vector strangeness of nucleons
realized in a baryon model

  \end{bf}

  \end{Large}

\vspace{4mm}

  \begin{Large}

\center{D. Horvati\'c$^{1}$, D. Klabu\v{c}ar$^{2,\rm a,b,c}$ and D. Mekterovi\'{c}$^{3}$  }

\vspace{6mm}

{\footnotesize

{\small $^1$ Department of Physics, Faculty of
Science, University of Zagreb, \\ Bijeni\v cka c. 32, Zagreb,
Croatia    }

\vspace{0.2cm}

{\small $^2$ International Centre for Theoretical Physics, Trieste, Italy }

\vspace{0.2cm}

%

$^3$ Rudjer Bo\v{s}kovi\'{c} Institute,
         P.O. Box 180, HR-10002 Zagreb, Croatia \\ }

\vspace{0.2cm}

\end{Large}

\end{center}

\vspace{0.1cm}
\begin{center}
  {\bf Abstract}
\end{center}
\begin{quotation}
\noindent

\noindent
Most of model considerations of the hidden nucleon strangeness,
as well as some preliminary experimental evidence, led to the
expectations of relatively sizeable strange vector form factors
of the proton. For example, it seemed that the contribution of 
the fluctuating strange quark-antiquark pairs accounts for as 
much as one tenth of the proton's magnetic moment. 
By the same token, baryon models which failed to produce 
the ``vector strangeness" of the nucleon seemed disfavored.
Recently, however, more accurate measurements and more 
sophisticated data analysis, as well as lattice simulations,
revealed that the form factors associated with the vector 
strangeness of the nucleon are much smaller than thought 
previously; in fact, due to the experimental uncertainties,
the measured strange vector-current proton form factors
may be consistent with zero. In the light of that, we 
re-asses the merit of the baryon models leading to little 
or no vector strangeness of the nucleon. It is done on the 
concrete example of the baryon model which essentially 
amounts to the MIT bag enriched by the diluted instanton 
liquid.

\vspace{0.2cm}


\end{quotation}

\vfill

----------------------------------------

$^{\rm a}$ Corresponding author, e-mail: klabucar@phy.hr

$^{\rm b}$ Senior associate of the Abdus Salam ICTP

$^{\rm c}$ Permanent address: Department of Physics, Faculty of
Science, \\ \indent University of Zagreb, Bijeni\v cka c. 32, Zagreb,
Croatia

\newpage

\markright{$$\small
Unexpectedly small empirical vector strangeness of nucleons
realized in a baryon model
$$}


\section{Introduction}
\label{INTRO}

The simple, ``naive" picture of hadrons, based on the models where 
only valence quarks are present, suffers a radical change when one 
takes into account the quantum field effects and consequently, the 
presence of the fluctuating virtual quark-antiquark ($q\bar q$) pairs. 
Indeed, the production of such fluctuating virtual pairs by 
interactions present in quantum chromodynamics (QCD) can be quite 
significant for the light quark flavors $q=u,d,s$.

The nucleon states, although they of course contain no {\it net} strangeness,
are thus expected to have also an intrinsic strangeness component due to 
fluctuating $s\bar s$ pairs. Precisely because nucleons have no valence 
strange ($s$) quarks, quantities originating from, or being influenced
by strange quarks, provide us with the information on the dynamics of 
virtual quarks within nucleons. 

A review of the issue of nucleon strangeness containing a very complete 
set of original references is for example Ref. \cite{28}, and for more
recent discussions of nucleon structure addressing also the nucleon 
strangeness issue, see for example 
Refs. \cite{Beck:2001yx,Ramsey-Musolf:2005rz,Thomas:2005qb}. 

Such considerations have led to the wide-spread belief that strange
quarks and antiquarks play a major role in protons and neutrons. 
For example, although estimates of such $s\bar s$ contributions to 
the nucleon mass vary between 100 to 300 MeV, they are in any case
very significant, between 10\% to 30\% of the nucleon mass.
However, the quantities related to the present considerations
are the proton magnetic moment and related electromagnetic form
factors of the proton, so now we turn to them.

\section{Strange form factors}
\label{FormFactors}

The Dirac and the Pauli strange vector form factors
(denoted by $F_{1}^{s}$ and $F_{2}^{s}$, respectively) of 
the nucleon ($N$) are defined through the matrix element of 
\begin{equation}
V_{\mu}^{s} = \bar{s}\gamma_{\mu}s \;,
\end{equation}
namely the vector current of $s$-quarks: 
\begin{equation}
\bra{N} V_{\mu}^{s} \ket{N} =
\bra{N}\bar{s}\gamma_{\mu}s\ket{N} = \bar{u}_{N}(p')
\left[ F_{1}^{s}(q^2)\gamma_{\mu} + F_{2}^{s}(q^2)
\frac{i\sigma_{\mu\nu} q^{\nu}} {2 M_{N}}\right]u_{N}(p) \;,
\label{f1f2}
\end{equation}
where $u_N$ is a nucleon spinor, $p'$ and $p$ are nucleon
momenta, and $q=p' - p$ is the transferred momentum.

Although $F_{1}^{s}(0)=0$, as it is the net nucleon strangeness, 
its momentum dependence determines the strangeness radius
\begin{equation}
r_{s}^{2}=6\frac{d}{dq^2}F_{1}^{s}(q^2)\Bigg|_{q^2=0} \;,
\end{equation}
while the strange magnetic moment is given by
\begin{equation}
\mu_{s}=F_{2}^{s}(0) \;.
\end{equation}

For comparison with experimental data, the (stran\-ge) Sachs form factors
$G^s_E$ (electric) and $G^s_M$ (magnetic) are widely used:
\begin{eqnarray}
G^s_E(q^2) &=& F^s_1(q^2) + \frac{q^2}{4 M_N^2} F^s_2(q^2) \;,\nonumber\\
G^s_M(q^2) &=& F^s_1(q^2) + F^s_2(q^2) \;.
\end{eqnarray}
Taking the non-relativistic nucleon spinor (of momentum $p$
and spin projection $\footnotesize\zeta$)
\begin{equation}
u_N(p,\zeta) = \sqrt{\frac{E+M_N}{2 E}} \left(
\begin{array}{c}
\chi_{\zeta} \\
\frac{\displaystyle{ \mbox{\boldmath$\sigma$} \cdot \vec{p}}}{
\displaystyle{E+m}} \chi_\zeta
\end{array}
\right)  \;,
\end{equation}
where $\chi_\zeta$ is a two-component Pauli spinor. 
Going to the Breit frame defined by
\begin{eqnarray}
q^{\mu} &=& (q^{0}, \vec{q} ) = (0,\vec{q}_B) \;, \nonumber\\
\vec{p} &=& \frac{ \vec{q}_{B}}{2}\; , \;
\vec{p}' = -\frac{ \vec{q}_{B}}{2} \;,
\end{eqnarray}
the components of the vector-current nucleon matrix elements
are expressed by Sachs form factors through the relations
\begin{eqnarray}
\langle N(p',\zeta')| V_0^s | N(p,\zeta)\rangle &=& \frac{m}{E}
\chi^{\dagger}_{\zeta'}\chi_\zeta G_{E}^{s}(-\vec{q}^2_B) \;, \\
\langle N(p',\zeta')| \vec{V}^s | N(p,\zeta)\rangle &=& \frac{1}{2 E}
\chi_{\zeta'}^{\dagger} i ( \mbox{\boldmath$\sigma$} \times \vec{q}_B)
\chi_\zeta G_{M}^{s}(-\vec{q}^2_B) \;.  
\label{defGm}
\end{eqnarray}

\section{How we get vector strangeness}
\label{Evaluating}

In the recent past, numerous model and lattice calculations gave 
very differing results for such strangeness contributions. For
example, various results on the $s\bar s$ contribution $\mu_s$ to
the proton magnetic moment range from 0.003 to as high as 0.8 
nucleon magnetons ($\mu_N$) in absolute magnitude. What is more,
they differ among each other even up to a sign. (For overview 
and references, see Ref. \cite{Beck:2001yx}.)

Overall, majority of the model calculations of nucleon strangeness
led to the expectations of substantial strangeness contributions to 
the vector form factors and the magnetic moment of the nucleon. 
In contrast to that, the model introduced by Klabu\v car {\it et 
al.} \cite{Klabucar:1998pb} and elaborated in Refs.
\cite{Klabucar:2000gu,Klabucar:2000rx}, yields 
zero results for these strange quantities, although it reveals 
substantial scalar strangeness. This is in accord with the 
conjecture \cite{Zh97} that a non-trivial QCD-vacuum structure 
selects the pseudoscalar and scalar channels, which experience 
the axial and trace anomaly, respectively. However, in the light 
of the aforementioned prevalence of model results indicating 
that the vector strangeness of the proton is probably significant,
the vanishing vector strangeness obtained in the model of
Refs. \cite{Klabucar:1998pb,Klabucar:2000gu} seemed as a 
drawback, a weakness of that model. Thus, when the model was
further developed \cite{diplomskiDM,Klabucar:2003mh}, the 
vector strangeness and its significance was not discussed,
it was hardly even mentioned. (Only the scalar strangeness for the 
improved model was discussed \cite{Klabucar:2003mh}, and at 
length.)

Both in Refs. \cite{Klabucar:1998pb,Klabucar:2000gu} and 
Refs. \cite{diplomskiDM,Klabucar:2003mh},
the model is essentially the MIT bag model enriched by the 
presence of a dilute instanton liquid
\cite{Klabucar:1998pb,Klabucar:2000gu,Klabucar:2000rx},
so that it focuses on QCD-vacuum fluctuations as given by
the instanton-liquid model \cite{ShVZ80,DiP84,NoVZ89}. 
However, 
Refs. \cite{Klabucar:1998pb,Klabucar:2000gu,Klabucar:2000rx} 
employed the so-called linearized approximation \cite{Klabucar:1994qf}, 
which implies freezing the baryon radii in their original MIT values.  
In Refs. \cite{diplomskiDM,Klabucar:2003mh} this approximation 
was removed: the baryon bag radii were allowed to vary 
in the course of parameter fitting to the masses of the baryons 
from the ground state octet and decuplet. (The re-fitting was 
performed so that the radii had to satisfy the pressure-balance 
condition \cite{diplomskiDM,Klabucar:2003mh}. For details of 
the re-fitting, see Ref. \cite{diplomskiDM}.)

\begin{figure}
\centerline{\resizebox{0.45\textwidth}{!}{\includegraphics{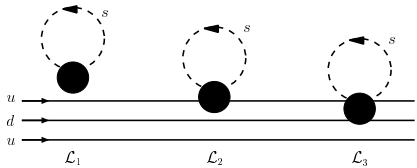}}}
\caption{Instanton-induced local strangeness induced in the proton by
the effective one-, two- and three-body operators in the interaction
(\ref{Linst}), namely $\calL_{1}$, $\calL_{2}$, and $\calL_{3}$,
respectively. Non-strange quarks are denoted by solid lines, and
strange ones by dashed lines.\label{figb}}
\end{figure}

In any case, in all Refs. 
\cite{Klabucar:1998pb,Klabucar:2000gu,Klabucar:2000rx,diplomskiDM,Klabucar:2003mh},
the instanton-induced interaction of the instanton-liquid 
model \cite{ShVZ80,DiP84,NoVZ89} produces QCD-vacuum fluctuations,
including presently interesting $s$-quark loops,
schematically shown in Fig. \ref{figb}.
Let us denote the corresponding Lagrangian density by $\calL_{I}$.
The instanton-induced interaction contains the one-, two-, and 
three-body operators (respectively denoted by $\calL_{1}$, 
$\calL_{2}$, and $\calL_{3}$ and given explicitly in 
Refs. \cite{Klabucar:1994qf,Klabucar:1998pb,Klabucar:2000gu}),
\begin{equation}
\calL_{I}=\calL_{1}+\calL_{2}+\calL_{3} \;.
\label{Linst}
\end{equation}

\begin{figure}
\centerline{\resizebox{0.40\textwidth}{!}{\includegraphics{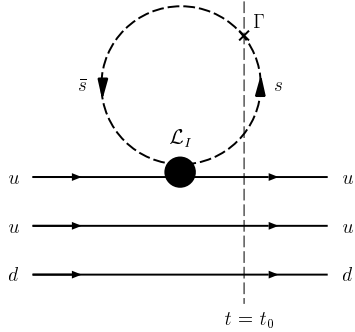}}}
\caption{A non-vanishing $s\bar s$ component of the nucleon state
found (at the moment $t=t_0$)
by the probe coupled at the vertex $\Gamma$ (denoted by $\times$).
More precisely, this graph is that part of the proton response which
arises only through one interaction ${\cal L}_I$. In the concrete case 
depicted in this figure, it is the two-body interaction $\calL_{2}$.
\label{strfig1}}
\end{figure}

Fig. \ref{strfig1} shows how an external probe couples at
the vertex $\Gamma$ to such an $s$-quark loop produced
by $\calL_{I}$. Various couplings are possible:
$\Gamma=1, \gamma_{5}, \gamma_{\mu}$, $\gamma_{\mu}\gamma_{5}, 
\sigma_{\mu\nu}$, corresponding, respectively, to the scalar,
pseudoscalar, vector, axial vector, and tensor strangeness.

The instanton-induced interaction (\ref{Linst}) contains 
the instanton density $n$ as an overall factor.
In the linearized approximation, the instanton density inside the 
bag was found \cite{Klabucar:1994qf,Klabucar:1998pb} to be very 
depleted with respect to its value in the nonperturbative QCD vacuum.
It is then a good approximation to keep only the first term in the 
perturbation series in the interaction $-\mathcal{L}_{I}$. Thus, the 
result of Ref. \cite{Klabucar:1998pb} for evaluating the 
nucleon-strangeness matrix element can be written as
\begin{eqnarray}
\lefteqn{\langle N | : \bar{s} \Gamma s : | N \rangle = 
 i \int^{\infty}_{-\infty}
dt' \; \langle N_0 | \hat{T} : \int d^3x } \nonumber \\
&\times& \bar{s}( \vec{x}, t_0) \Gamma s ( \vec{x}, t_0):
  \; :\! \int d^3 y\: \mathcal{L}_{I}
( \vec{y}, t'): | N_0 \rangle \;,
\label{inststran}
\end{eqnarray}
where $:...:$ denotes the normal ordering, and $| N_0 \rangle$
is the model nucleon ground state composed of the non-strange 
valence quarks only. Note that $s(\vec{x},t)$ denotes the 
strange quark field. In our model calculation, for any quark
flavor $q$ (and we are now interested only in the light flavors,
$q=u,d,s$), we expand the quark fields $q(\vec{x}, t)$ in 
an appropriate wave-function basis $\{ q_{K} \}$
in terms of creation
(${\cal U}^\dagger_{K}, {\cal D}^\dagger_{K}, {\cal S}^\dagger_{K}$)
and annihilation (${\cal U}_{K}, {\cal D}_{K},{\cal S}_{K}$) 
operators of {\it dressed} quarks and antiquarks:
\begin{equation}
q(\vec{x}, t)=\sum_K \left[ \calQ_K\, q_{K}({\vec{x}})e^{-i\omega_K t} +
  \calQ^{c^\dagger}_K\, q_{K}^{c} ({\vec{x}})e^{ i\omega_K t}
        \right] \;.
\label{expandq}
\end{equation}
Here, $q_{K}(\vec{r})$ denotes a model wave function of a quark 
of flavour $q$, where $K$ stands for the set of quantum numbers 
labeling a model quark state.

As already said, for concrete evaluations of a nucleon-strangeness 
matrix element (\ref{inststran}), Refs.
\cite{Klabucar:1998pb,Klabucar:2000gu,Klabucar:2000rx,diplomskiDM,Klabucar:2003mh}
chose to employ the MIT bag model. With this choice, $q_{K}(\vec{r})$ 
is the solution for the quark in the $K$-th mode of the MIT bag.
For the wavefunctions and other details of the model calculations, 
see especially Ref. \cite{Klabucar:2000gu}.

For a concrete evaluation of a nucleon-strangeness matrix element 
(\ref{inststran}), one should also specify the pertinent tensor 
structure of the vertex $\Gamma$ -- i.e., which kind of nucleon 
strangeness one wants to evaluate.  The presently relevant case is 
$\Gamma=\gamma_{\mu}$, i.e., the case of the ``vector strangeness"
of the proton, as we are interested in the form factors of the 
vector current -- see Eq. (\ref{f1f2}).

In order to calculate the contribution of the instanton-induced vector strange 
current inside the MIT bag, we have to identify the form factors in 
(\ref{defGm}) with the Fourier-transformed vector current within the bag:
\begin{eqnarray}
\lefteqn{\langle N(p')|: V_{\mu}^{s}:| N(p) 
\rangle  } \nonumber \\
& = & \langle N(p')|: \! \int d^3 r\: e^{-i \vec{q}_B\cdot 
\vec{r} } \bar{s}(\vec{r})
\gamma_{\mu} s(\vec{r}): |N(p)\rangle,
\end{eqnarray}
using the static limit $q \rightarrow 0$. From the $V^s_0$
component of the vector current, the electric Sachs form factor
$G_{E}^{s}(q^2=0)= 0$ (in the leading order) 
in the original model employing linearized approximation
\cite{Klabucar:1998pb,Klabucar:2000gu,Klabucar:2000rx},
since $\langle N(p')|: V_{0}^{s}:| N(p)\rangle$ evaluated 
through Eq. (\ref{inststran}) vanishes identically in the 
original model. 
Also, the calculation for the space components $\vec{V}^s$ 
yields the vanishing magnetic form factor, $G_M^s(0) = 0$.
This implies the vanishing strange magnetic moment
\begin{equation}
\mu_{s}=F_{2}^{s}(0) = 0 \;,
\label{mus}
\end{equation}

Now we want to point out that this vanishing of 
$G_{E}^{s}(q^2=0)$ still holds in the improved 
version of the model \cite{diplomskiDM,Klabucar:2003mh} 
not employing linearized approximation, which can be 
seen easily in the evaluation of Eq. (\ref{inststran}). 
Obtaining the vanishing of $G_M^s(0)$ or, equivalently, 
$\mu_{s}=0$, is not so trivial in the explicit calculation, 
as it happens due to
a subtle cancellation among the contributions of quarks
in the loop with different spin orientations (and the
calculation requires careful handling of mode sums 
resulting from Eqs. (\ref{inststran}) and (\ref{expandq})
-- see. Eq. (28) in Ref. \cite{Klabucar:1998pb}). In any 
case, one can get some non-vanishing vector strangeness 
only from higher orders in the instanton-induced interaction.
However, such contributions would be very small and could 
be neglected in deriving Eq. (\ref{inststran}) if the 
instanton density allowed inside the bag is sufficiently 
low, and this was certainly the case in the linearized 
approximation \cite{Klabucar:1998pb}.
 
In retrospect, one should note that such a result of the 
explicit model evaluation is expected in any model on general
grounds, since there is the qualitative mechanism of the 
suppression of the violation of the Okubo-Zweig-Iizuka 
rule in the vector channel \cite{Geshkenbein:1979vb}.
It is due to the spin structure in the 'tHooft's single
instanton-induced quark interaction.
On the model level, 
removing the linearized approximation
amounts (in the sense of implications on re-fitting
the bag model parameters) to allowing the bag radii
to vary freely, which cannot upset the
aforementioned cancellation in Eq. (\ref{inststran}). 
However, removing the
linearized approximation also led \cite{Klabucar:2003mh}
to much larger values of the instanton density inside
the MIT bag than before, in Ref. \cite{Klabucar:1998pb}).
The question then arises whether Eq. (\ref{inststran})
remains a good approximation, i.e., whether one can 
still consider the second and higher order terms in 
the instanton-induced interaction as negligibly small.


Without the linearized approximation, 
Ref. \cite{Klabucar:2003mh} obtained solutions where the 
densities $n$ inside the bag are an order of magnitude 
larger than in the linearized approximation, where it 
was found \cite{Klabucar:1998pb} to be just 
$n = 0.266 \cdot 10^{-4}$ GeV$^{4}$.  Nevertheless, for all 
acceptable phenomenological fits without the linearized 
approximation, Ref. \cite{Klabucar:2003mh} found that 
instanton densities possible inside the MIT bag, are still 
appreciably lower (at least by the factor of 3 or more) 
than $n_0$, the usual non-perturbative vacuum instanton 
density in the non-perturbative vacuum, where 
$n_0 \approx 1 $ fm$^{-4}$ $= 1.6 \cdot 10^{-−3}$ GeV$^{4}$.
Thus, is still justified to neglect the higher order instanton 
contributions and adopt the first order approximation 
(\ref{inststran}).
(It should be noted that Ref. \cite{Kochelev:1985de} 
also estimated it could neglect higher orders in the 
instanton-induced interaction, although it used the 
full, non-depleted value of the instanton density, 
i.e., $n_0$, the instanton density appropriate for the 
non-perturbative QCD vacuum, in a part of the bag volume.)

\section{Discussion and conclusion}
\label{Conclusion}

At the time of publication of Ref. \cite{Klabucar:1998pb},
such results on vector strangeness seemed compatible with
the experimental results \cite{Mu97,An99} available then.
However, since that time, not only other theoretical 
considerations, but, more importantly, preliminary 
announcements of more precise experimental results seemed, 
for a while, to point out that vector strangeness is rather 
large and that our approach is not suitable for treating it.
The strange form factors and magnetic moment were therefore 
not considered in the improved version of the model beyond the 
linearized approximation \cite{diplomskiDM,Klabucar:2003mh}.
Such situation with the strange vector form factors seemed
confirmed when the G0 collaboration, performing high-precision
measurement at Jefferson Lab, announced large 
positive results for the magnetic form factor (over substantial
range of momentum transfers, $0.12\leq Q^2 \leq 1.0$ GeV$^2$)
\cite{Armstrong:2005hs}.

More recent developments, however, took a surprising turn.
One may first note the recent lattice results which 
differ from the quoted G0 results even by the sign
($G_M^s = (-0.046 \pm 0.022) \mu_N$ 
\cite{Leinweber:2004tc,Thomas:2006jf}). 
The most notable are of course the experimental results of the 
nucleon strange form factors,
also obtained at Jefferson Lab but by HAPPEX collaboration,
which show that the electric form factor essentially 
vanishes: $G_E^s(Q^2=0.1\, {\rm GeV}^2) = -0.01 \pm 0.03$
\cite{Aniol:2005zf,Aniol:2005zg}. This is in excellent 
agreement \cite{Thomas:2006jf} with the lattice results 
\cite{Leinweber:2006ug}
also essentially showing the vanishing of the same quantity, 
obtained by the same method as $G_M^s$ \cite{Leinweber:2004tc}. 
Careful analyses of the methods of extracting individual
form factors revealed that it was difficult to perform an 
experimental separation of the individual form factors, and 
that it was not always clear what had been measured and what 
the role of theoretical input had been \cite{Thomas:2006jf}.
The proper insight has finally been gained by unifying all
pertinent world data, which means the results of SAMPLE 
\cite{Spayde:2003nr}, A4 \cite{Maas:2004ta,A4nemaNaQspires},
G0 \cite{Armstrong:2005hs} and HAPPEX 
\cite{Aniol:2005zf,Aniol:2005zg} collaborations,
 and by {\it joint analysis} of various form
factors. For our present purposes, the most illustrative
is Fig. 2 from Ref. \cite{Aniol:2005zg}, showing the 
data on $G_E^s$ and $G_M^s$ from SAMPLE, A4, G0 and HAPPEX
collaborations (along with some theoretical predictions).
In that plot, the ellipse shows the 95\% confidence level
for the possible values of $G_E^s$ and $G_M^s$ and indicates 
that the vector strangeness is not that large as people 
came to think previously. The best fit values are 
$G_E^s = -0.01 \pm 0.03$, which is perfectly consistent
with zero, and $G_M^s = (+0.55 \pm 0.28) \mu_N$. 
While this fit thus favors nonzero values for $G_M^s$, 
we should note {\it i)} the
suspicious sign difference with respect to the lattice 
results for $G_M^s$ \cite{Leinweber:2004tc,Thomas:2006jf},
and {\it ii)} that the value  $G_M^s = 0$ is still allowed
at the 95\% confidence level.

In conclusion, we have shown how the improved version
\cite{diplomskiDM,Klabucar:2003mh} of the model 
\cite{Klabucar:1998pb,Klabucar:2000gu,Klabucar:2000rx}
which we used to study various aspects of the hidden 
nucleon strangeness, also yields the zero vector
strangeness of the nucleon, namely the vanishing 
form factors $G_E^s$ and $G_M^s$ of the nucleons. 
While until recently this was considered wrong and 
an unpleasant artefact of the model, more precise 
measurements and more sophisticated data analysis,
along with lattice QCD simulations, now show that 
such a vanishing vector strangeness may well be 
genuine, or at least that it is a good approximation.
This simple model in the both variants
\cite{Klabucar:1998pb,Klabucar:2000gu,Klabucar:2000rx,diplomskiDM,Klabucar:2003mh}
in the end turned out to be more physical than many
very sophisticated models designed to produce
a large vector strangeness of the nucleon.


\section*{Acknowledgments}
\noindent
D. Klabu\v car acknowledges the hospitality of the Abdus Salam ICTP in Trieste.


\newpage
\begin{thebibliography}{99}

\bibitem{28} R. Decker, M. Nowakowski, U. Wiedner,
                       Fort. Phys. {\bf 41}, 87 (1993) 87.


\bibitem{Beck:2001yx}
  D.~H.~Beck and R.~D.~McKeown,
  Ann.\ Rev.\ Nucl.\ Part.\ Sci.\  {\bf 51} (2001) 189.



\bibitem{Ramsey-Musolf:2005rz}
  M.~J.~Ramsey-Musolf,
  Eur.\ Phys.\ J.\ A {\bf 24S2} (2005) 197.


\bibitem{Thomas:2005qb}
  A.~W.~Thomas, R.~D.~Young and D.~B.~Leinweber,
  arXiv:nucl-th/0509082.


\bibitem{Klabucar:1998pb}
  D.~Klabu\v car, K.~Kumeri\v cki, B.~Meli\'c and I.~Picek,
  Eur.\ Phys.\ J.\  C {\bf 9} (1999) 589.

\bibitem{Klabucar:2000gu}
  D.~Klabu\v car, K.~Kumeri\v cki, B.~Meli\'c and I.~Picek,
  Fizika B {\bf 8} (1999) 505.

\bibitem{Klabucar:2000rx}
  D.~Klabu\v car, K.~Kumeri\v cki, I.~Picek and B.~Meli\'c,
  Czech.\ J.\ Phys.\  {\bf 50S1} (2000) 187.

\bibitem{Zh97}
A.~R. Zhitnitsky, Phys. Rev. {\bf D55} (1997) 3006.


\bibitem{diplomskiDM}
D. Mekterovi\'c,
``Instantons and baryon mass spectrum in the MIT bag model",
diploma thesis (in Croatian), Physics Department, Faculty of Science, 
University of Zagreb, August 2001, thesis advisor Dubravko Klabu\v car.


\bibitem{Klabucar:2003mh}
  D.~Klabu\v car, K.~Kumeri\v cki, D.~Mekterovi\'c and B.~Podobnik,
  Eur.\ Phys.\ J.\  C {\bf 29} (2003) 71.


\bibitem{ShVZ80}
M.~A. Shifman, A.~I. Vainshtein and V.~I. Zakharov, 
                    Nucl. Phys. {\bf B163} (1980) 46.


\bibitem{DiP84}
D.~I. Diakonov and V.~Y. Petrov, Nucl. Phys. {\bf B245} (1984) 259.


\bibitem{NoVZ89}
M.~A. Nowak, J.~J.~M. Verbaarschot and I.~Zahed, 
                     Nucl. Phys. {\bf B324} (1989) 1.


\bibitem{Klabucar:1994qf}
  D.~Klabu\v car,
  Phys.\ Rev.\  D {\bf 49} (1994) 1506.


\bibitem{Geshkenbein:1979vb}
  B.~V.~Geshkenbein and B.~L.~Ioffe,
  Nucl.\ Phys.\  B {\bf 166} (1980) 340.


\bibitem{Kochelev:1985de}
  N.~I.~Kochelev,
  Sov.\ J.\ Nucl.\ Phys.\  {\bf 41} (1985) 291
  [Yad.\ Fiz.\  {\bf 41} (1985) 456].


\bibitem{Mu97}
B.~Mueller et~al., Phys. Rev. Lett. {\bf 78} (1997) 3824.

\bibitem{An99}
K.~A. Aniol et~al., Phys. Rev. Lett. {\bf 82} (1999) 1096.


\bibitem{Armstrong:2005hs}
  D.~S.~Armstrong {\it et al.}  [G0 Collaboration],
  Phys.\ Rev.\ Lett.\  {\bf 95} (2005) 092001.


\bibitem{Leinweber:2004tc}
  D.~B.~Leinweber {\it et al.},
  Phys.\ Rev.\ Lett.\  {\bf 94} (2005) 212001.


\bibitem{Thomas:2006jf}
  A.~W.~Thomas and R.~D.~Young,
  Nucl.\ Phys.\  A {\bf 782} (2007) 1.

\bibitem{Aniol:2005zf}
  K.~A.~Aniol {\it et al.}  [HAPPEX Collaboration],
  Phys.\ Rev.\ Lett.\  {\bf 96} (2006) 022003.


\bibitem{Aniol:2005zg}
  K.~A.~Aniol {\it et al.}  [HAPPEX Collaboration],
  Phys.\ Lett.\ B {\bf 635} (2006) 275.



\bibitem{Leinweber:2006ug}
  D.~B.~Leinweber {\it et al.},
  Phys.\ Rev.\ Lett.\  {\bf 97} (2006) 022001.


\bibitem{Spayde:2003nr}
  D.~T.~Spayde {\it et al.}  [SAMPLE Collaboration],
  Phys.\ Lett.\  B {\bf 583} (2004) 79.

\bibitem{Maas:2004ta}
  F.~E.~Maas {\it et al.}  [A4 Collaboration],
  Phys.\ Rev.\ Lett.\  {\bf 93} (2004) 022002.

\bibitem{A4nemaNaQspires}
F.~E.~Maas {\it et al.}  [A4 Collaboration], 
Phys.\ Rev.\ Lett.\  {\bf 94} (2005) 152001.

\end{thebibliography}
\end{document}